\documentclass[aps,prl,superscriptaddress,floatfix,reprint,showkeys,showpacs]{revtex4-1}

\usepackage{amsmath}
\usepackage{graphicx}

\begin{document}
\title{Complex Squeezing and Force Measurement Beyond the Standard Quantum Limit}

\author{L.~F.~Buchmann}
\email{lbuchmann@phys.au.dk}
\affiliation{Department of Physics and Astronomy, Aarhus University, Ny Munkegade 120, DK 8000 Aarhus C, Denmark}
\affiliation{Department of Physics, University of California, Berkeley, California 94720, USA}

\author{S.~Schreppler}
\affiliation{Department of Physics, University of California, Berkeley, California 94720, USA}
\author{J.~Kohler}
\affiliation{Department of Physics, University of California, Berkeley, California 94720, USA}
\author{N.~Spethmann}
\affiliation{Department of Physics, University of California, Berkeley, California 94720, USA}
\affiliation{Fachbereich Physik, Technische Universit\"at Kaiserslautern, 67663 Kaiserslautern, Germany}
\author{D.~M.~Stamper-Kurn}
\affiliation{Department of Physics, University of California, Berkeley, California 94720, USA}
\affiliation{Materials Sciences Division, Lawrence Berkeley National Laboratory, Berkeley, California 94720, USA}
\begin{abstract}
A continuous quantum field, such as a propagating beam of light, may be characterized by a squeezing spectrum that is inhomogeneous in frequency.  We point out that homodyne detectors, which are commonly employed to detect quantum squeezing, are blind to squeezing spectra in which the correlation between amplitude and phase fluctuations is complex.  We find theoretically that such complex squeezing is a component of ponderomotive squeezing of light through cavity optomechanics.  We propose a detection scheme, called synodyne detection, which reveals complex squeezing and allows the accounting of measurement back-action. Even with the optomechanical system subject to continuous measurement, such detection allows the measurement of one component of an external force with sensitivity only limited by the mechanical oscillator's thermal occupation. 
\end{abstract}
\maketitle

\newcommand{\dmskcomment}[1]{\textbf{XXX #1 XXX}}
The non-trivial commutation relations of quantum operators lead to minimum uncertainty relations, measurement back-action and the standard quantum limit (SQL) to measurement precision \cite{RMP}. To beat such a limit, one may use squeezed states of quantum fields \cite{oldsqueezing1, oldsqueezing2}, which may involve optical fields~\cite{squeezexp}, mechanical oscillators~\cite{mechsqueezing1, mechsqueezing2}, spins~\cite{spinsqueezing}, atomic numbers~\cite{numbersqueezing} and other observables.

The simplest treatment of squeezing involves a single-mode quantum field~\cite{qoptics}, for which one defines two orthogonal field quadratures: the observables $\hat{X}_{\mathrm{AM}} = (\hat{a}^\dagger + \hat{a})/\sqrt{2}$ and $\hat{X}_{\mathrm{PM}} = i(\hat{a}^\dagger - \hat{a})/\sqrt{2}$, where $\hat{a}$ is a dimensionless bosonic destruction operator. These observables describe amplitude (AM) or phase fluctuations (PM) atop a state with a real expectation value, typically a coherent state at some carrier frequency $\omega_0$.  A quantum state is squeezed if one can identify a quadrature
\begin{equation}
\hat{X}_\varphi = \cos(\varphi) \hat{X}_{\mathrm{AM}} + \sin(\varphi) \hat{X}_{\mathrm{PM}}
\end{equation}
for which the variance is below that of a vacuum state, i.e. $\langle\hat{X}_\varphi^2\rangle - \langle \hat{X}_\varphi\rangle^2<1/2$.
To detect such squeezing it is sufficient to use a homodyne receiver, in which the quantum field is combined linearly with a coherent-state local oscillator (LO) with amplitude $\alpha = |\alpha| e^{i \theta-i\omega_0 t}$, and the two outputs of the combiner are detected with a square-law detector, for instance a photodetector for optical fields.  Choosing $\theta = \varphi$ causes the detector's optical shot noise (quantified by repeated measurement of identically prepared fields) to fall below the noise observed in detecting the vacuum field.

Nonlinear processes that produce squeezed light -- such as parametric down-conversion~\cite{paramsqueezing}, resonance fluorescence~\cite{resfluorescence}, and optomechanical interactions \cite{ponderomotiveAtom,ponderomotiveCrystal,ponderomotiveMembrane} -- are characterized by finite response times, necessitating a multimode approach and leading to an inhomogeneous squeezing spectrum \cite{JOSA}. Again, squeezing is conventionally regarded as being measured with a homodyne receiver, which performs a continuous measurement of the observable \mbox{$\hat{X}_{\theta}(t) = \cos(\theta) \hat{X}_{\mathrm{AM}}(t) + \sin(\theta)\hat{X}_{\mathrm{PM}}(t)$}, with the evolution at the carrier frequency $\omega_0$ absorbed into the definitions of the operators. A squeezed state is defined as one for which the noise spectral density \mbox{$S^\mathrm{hom}_\theta(\omega)\delta(\omega-\omega')=\frac{1}{2}\left\langle\{\tilde{X}_\theta(\omega),\tilde{X}_\theta(-\omega')\}\right\rangle$} falls below its value when detecting the vacuum state. Here $\tilde{X}_\theta(\omega)$ denotes the Fourier transform of the operator $\hat{X}_\theta(t)$, from which we have subtracted the expectation value. The level and quadrature angle of squeezing generally vary with detection frequency $\omega$.

In this letter, we illustrate that the restriction to standard homodyne detection overlooks a broader class of squeezed states, characterized by unequal-time correlations between field quadratures.  We demonstrate that these correlations are produced naturally by nonlinear optical systems, using the ponderomotive squeezing of light through cavity optomechanics as a specific example \cite{ponderomotiveAtom, ponderomotiveCrystal, ponderomotiveMembrane,pmTheory1, pmTheory2}.  This new \emph{complex squeezing} can be detected by mixing the propagating field with a suitably chosen LO, a method we call \emph{synodyne detection}. The resulting signal is truly squeezed, in that its noise level is lower than that obtained when detecting the vacuum state, and leads to improvements in measurement precision.

The real and imaginary parts of the Fourier transform of a detected homodyne signal at frequency $\omega\neq 0$ separately yield a measurement of two independent observables. These observables are time-weighted measurements of the quadrature operators and may be written as
\begin{eqnarray}\label{weightedops}
\tilde{X}^\xi_\varphi(\omega)& =& \int dt\, \hat{X}_\varphi(t) \cos(\omega t-\xi)
\end{eqnarray}
with $\varphi = 0 (\pi/2)$ for AM(PM) fluctuations of the field, and $\xi$ denoting the temporal phase, with $\xi=0(\pi/2)$ measured by the real (imaginary) part of the Fourier transform. The operator pair $\tilde{X}^0_{\mathrm{AM}}$ and $\tilde{X}^0_{\mathrm{PM}}$ (and, similarly, $\tilde{X}^{\pi/2}_{\mathrm{AM}}$ and $\tilde{X}^{\pi/2}_{\mathrm{PM}}$) measure non-commuting fluctuations of the field with equal strength at identical measurement times. Consequently, their commutator is nonzero.   In contrast, $\tilde{X}^0_\varphi$ and $\tilde{X}^{\pi/2}_{\varphi^\prime}$ commute for any choice of $\varphi, \varphi^\prime$ since they measure field fluctuations at different times.

Correlations between observables, which underlie squeezing, can occur between any two of these four observables ($\varphi\in\{0,\pi/2\}$ and $\xi\in\{0,\pi/2\}$). Standard homodyne detection does not offer access to all possible pairs.  For any detection angle, the variance of the real part of the Fourier transform of the measured time-record involves correlations $\langle\tilde{X}_{\mathrm{AM}}^{0}(\omega)\tilde{X}_{\mathrm{PM}}^{0}(-\omega)\rangle$ and the correlations appearing in the variance of the imaginary part are $\langle\tilde{X}_{\mathrm{AM}}^{\pi/2}(\omega)\tilde{X}_{\mathrm{PM}}^{\pi/2}(-\omega)\rangle$.  Such correlations may lead to noise reduction below the vacuum level.  However, correlations that may exist between AM fluctuations at cosine times ($\xi=0$) and PM fluctuations at sine times ($\xi=\pi/2$) cannot reduce the noise in such a detector.   The commutation relations among the four observables identified here are identical to those of two independent harmonic oscillators.  The latter form of correlations is akin to two-mode squeezing, i.e.\ EPR-type correlations \cite{eins35} between the motions of two objects.

To detect and exploit such unequal-time correlations, we require a ``generalized homodyne detector,'' which effectively measures the quantity
\begin{equation}
\tilde{X}_{\mu,\nu}(\omega) = \mu \tilde{X}_{\mathrm{AM}}(\omega) + \nu \tilde{X}_{\mathrm{PM}}(\omega),
\end{equation}
where $\mu$ and $\nu$ form a normalized spinor. In homodyne detection, we are restricted to choosing both $\mu$ and $\nu$ to be real (a common complex phase is eliminated simply by a common translation in time).  To reveal unequal-time correlations between observables, one would allow for a phase difference between $\mu$ and $\nu$, such that one of these components is necessarily complex, hence the name \emph{complex squeezing}.

It remains to show that such complex squeezing can be produced, that it can be detected, and that its exploitation can improve the sensitivity of a measurement. To illustrate these facts, we turn to the specific example of cavity optomechanics, the ponderomotive squeezing of light, and the task of force detection with an optomechanical sensor~\cite{forcesen}.  We emphasize that the concepts presented in this work are general, and should find application to other instances of squeezing in continuous quantum systems and their metrological applications.

\begin{figure}[t]
	\begin{center}
		\includegraphics[width=0.4\textwidth]{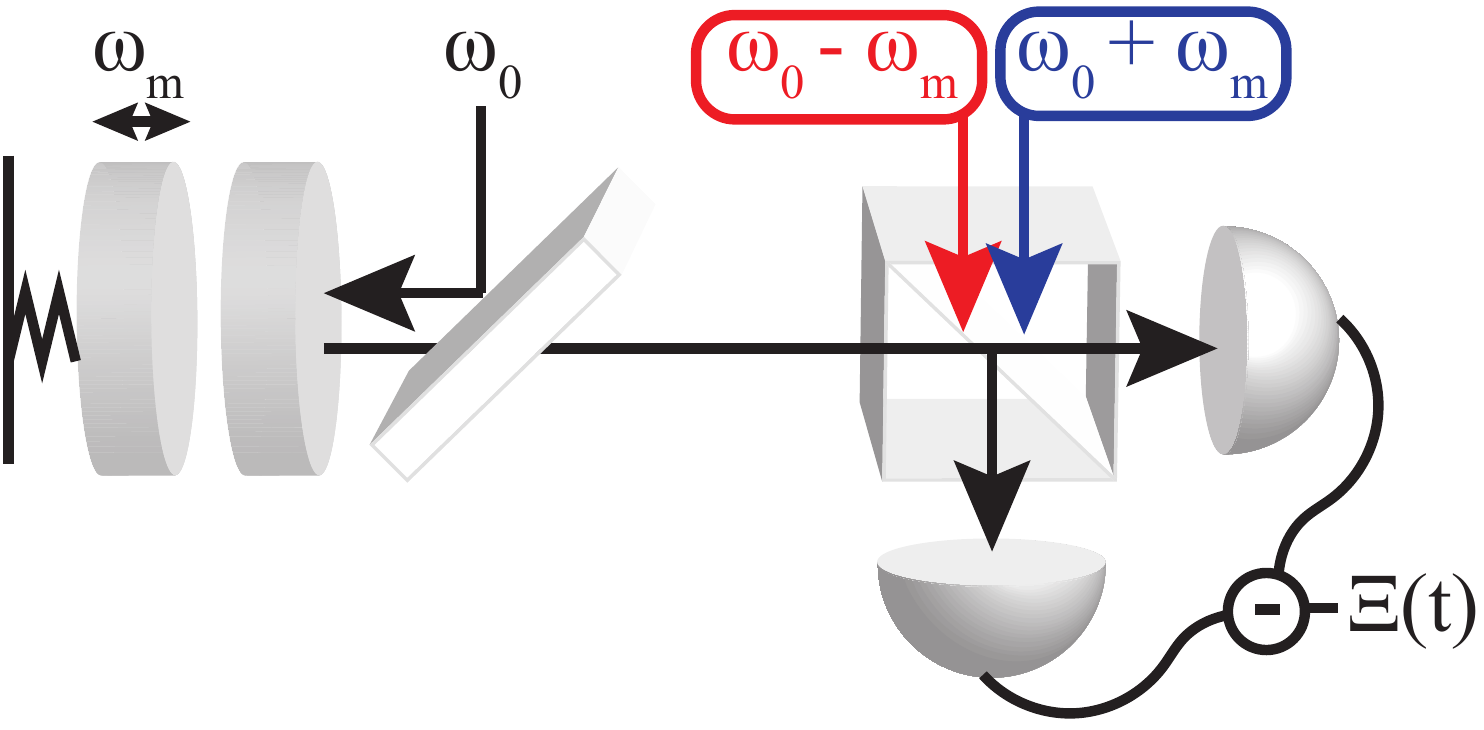}
		\caption{(Color online) The complex ponderomotive optical squeezing spectrum produced by cavity optomechanics is sensed by synodyne detection.  A cavity optomechanical system, consisting of an optical cavity and mechanical oscillator (at left) is driven by a coherent pump at optical frequency $\omega_0$.  The cavity emission exhibits a complex squeezing spectrum.  The squeezing is revealed by a synodyne detector, which uses a two-tone LO that is matched to measure one temporal phase of the complex squeezing spectrum, here sensed at the frequency $\omega_m$, as the dc signal in $\Xi_1(t)$. The setup also serves as a phase-sensitive force detector with perfect back-action accounting.}
		\label{scheme}
	\end{center}
\end{figure}

Consider a cavity optomechanical system in which displacement of a mechanical oscillator is dispersively coupled to a single electromagnetic mode of a one-sided cavity (Fig.\ \ref{scheme})~\cite{reviews}.   The cavity is pumped with an optical tone at frequency $\omega_0$.  Setting $\hbar=1$, and after linearization about the pump's coherent-state amplitude in a frame co-rotating with the pump light frequency, the system Hamiltonian is
\begin{align}
\mathcal{H}=\Delta\hat{a}^\dag\hat{a}+\omega_m\hat{b}^\dag\hat{b}+\frac{G}{\sqrt{2}}(\hat{a}+\hat{a}^\dag)(\hat{b}+\hat{b}^\dag)+\mathcal{H}_\mathrm{ext},
\end{align}
with mechanical resonance frequency $\omega_m$, the detuning of the pump from cavity resonance $\Delta$ and the dressed optomechanical coupling strength $G$~\cite{supp}.  The operators $\hat{a}$ and $\hat{b}$ refer to photon and phonon states, respectively.   Coupling to independent noise reservoirs of the optical and mechanical systems with energy decay rates $\kappa$ and $\gamma_m$, respectively, and also the effect of an external classical force, are described by $\mathcal{H}_\mathrm{ext}$.

When the cavity is pumped at its resonance frequency ($\Delta =0$), the PM quadrature of its output contains information about the displacement of the mechanical oscillator. We obtain the optical output from the standard input/output relation~\cite{qoptics}, $\hat{a}_\mathrm{out}(t)=\sqrt{\kappa}\hat{a}(t)-\hat{a}_\mathrm{in}(t)$, where $\hat{a}_\mathrm{in}$ is the vacuum input noise operator for which  $\langle\hat{a}_\mathrm{in}(t)\hat{a}^\dag_\mathrm{in}(t')\rangle=\delta(t-t')$. The Heisenberg equations of motion are easily inverted to give the output quadratures $\tilde{X}_\mathrm{AM}^\mathrm{out}$ and $\tilde{X}_\mathrm{PM}^\mathrm{out}$ in terms of the optical input noise in frequency space

\begin{subequations}\label{outputop}
\begin{align}
\tilde{X}_{\mathrm{AM}}^\mathrm{out}(\omega)=&\tilde{X}_{\mathrm{AM}}^\mathrm{in}(\omega),\\
\tilde{X}_{\mathrm{PM}}^\mathrm{out}(\omega)=&\tilde{X}_{\mathrm{PM}}^\mathrm{in}(\omega)+\chi_{BA}(\omega)\tilde{X}_{\mathrm{AM}}^\mathrm{in}(\omega)
\nonumber\\
&
+T_q(\omega)\tilde{Q}_\mathrm{in}(\omega)
+T_p(\omega)\tilde{P}_\mathrm{in}(\omega),\label{pmoutput}
\end{align}
\end{subequations}
where we absorbed the cavity time delay \mbox{$\chi_o=(\kappa/2-i\omega)/(\kappa/2+i\omega)$} into the definition of the optical input noise~\cite{supp}. The mechanical input operators satisfy $\langle\tilde{Q}_\mathrm{in}(\omega)\tilde{Q}_\mathrm{in}(\omega')\rangle=\langle\tilde{P}_\mathrm{in}(\omega)\tilde{P}_\mathrm{in}(\omega')\rangle=(\bar{\nu}+1/2)\delta(\omega+\omega')$ and $\langle\tilde{Q}_\mathrm{in}(\omega)\tilde{P}_\mathrm{in}(\omega')\rangle=-\langle\tilde{P}_\mathrm{in}(\omega')\tilde{Q}_\mathrm{in}(\omega)\rangle=i\delta(\omega+\omega')/2$ with $\bar{\nu}$ the thermal phonon occupation. The functions $T_i(\omega)$ describe the transduction of the mechanical signal unto the optical system
\begin{subequations}\label{transduction}
\begin{align}
T_q(\omega)=&G\sqrt{\kappa\gamma_m}\frac{\gamma_m/2+i\omega}{\kappa/2+i\omega}\chi_m(\omega),\\
T_p(\omega)=&G\sqrt{\kappa\gamma_m}\frac{\omega_m}{\kappa/2+i\omega}\chi_m(\omega),
\end{align}
\end{subequations}
with the mechanical susceptibility
\begin{align}
\chi_m(\omega)=\frac{1}{\gamma_m^2/4+\omega_m^2-\omega^2+i\omega\gamma_m},
\end{align}
and $\chi_{BA}(\omega)$ describes the back-action of the force measurement~\cite{quantMeas, MurchBA, PurdyBA, SuhBA, WilsonBA, TeufelBA}
\begin{align}\label{backaction}
\chi_{BA}(\omega)=\frac{G^2\kappa\omega_m}{\kappa^2/4+\omega^2}\chi_m(\omega).
\end{align}
The output light is completely characterized by the Hermitian covariance matrix \mbox{$[\mathcal{C}(\omega)]_{ij}\delta(\omega-\omega') = \frac{1}{2}\left\langle\left\{ \tilde{X}_i^\mathrm{out}(\omega), \tilde{X}_j^\mathrm{out}(-\omega')\right\}\right\rangle$}, where the indices 1 and 2 denote $\mathrm{AM}$ and $\mathrm{PM}$ respectively.

Homodyne detection of the cavity output light reveals the real part of the covariance $\left[\mathcal{C}(\omega)\right]_{12}$, as seen in the noise spectral density of such a measurement:
\begin{align}\label{homodynepsd}
S_\theta^\mathrm{hom}(\omega)=&\cos^2(\theta)[\mathcal{C}(\omega)]_{11}+\sin^2(\theta)[\mathcal{C}(\omega)]_{22}
\nonumber\\
&+2\sin(\theta)\cos(\theta)\mathrm{Re}\left([\mathcal{C}(\omega)]_{12}\right).
\end{align}
From Eqs.\ (\ref{outputop}) and (\ref{backaction}), we observe that equal-time correlations quantified by $\mathrm{Re}\left([\mathcal{C}(\omega)]_{12}\right)$ derive from the real part of the mechanical susceptibility $\chi_m(\omega)$.  The resulting ponderomotive squeezing has been observed on several optomechanical platforms \cite{ponderomotiveAtom,ponderomotiveCrystal,ponderomotiveMembrane} using standard detection techniques.

\begin{figure}[t]
\includegraphics[width=0.5\textwidth]{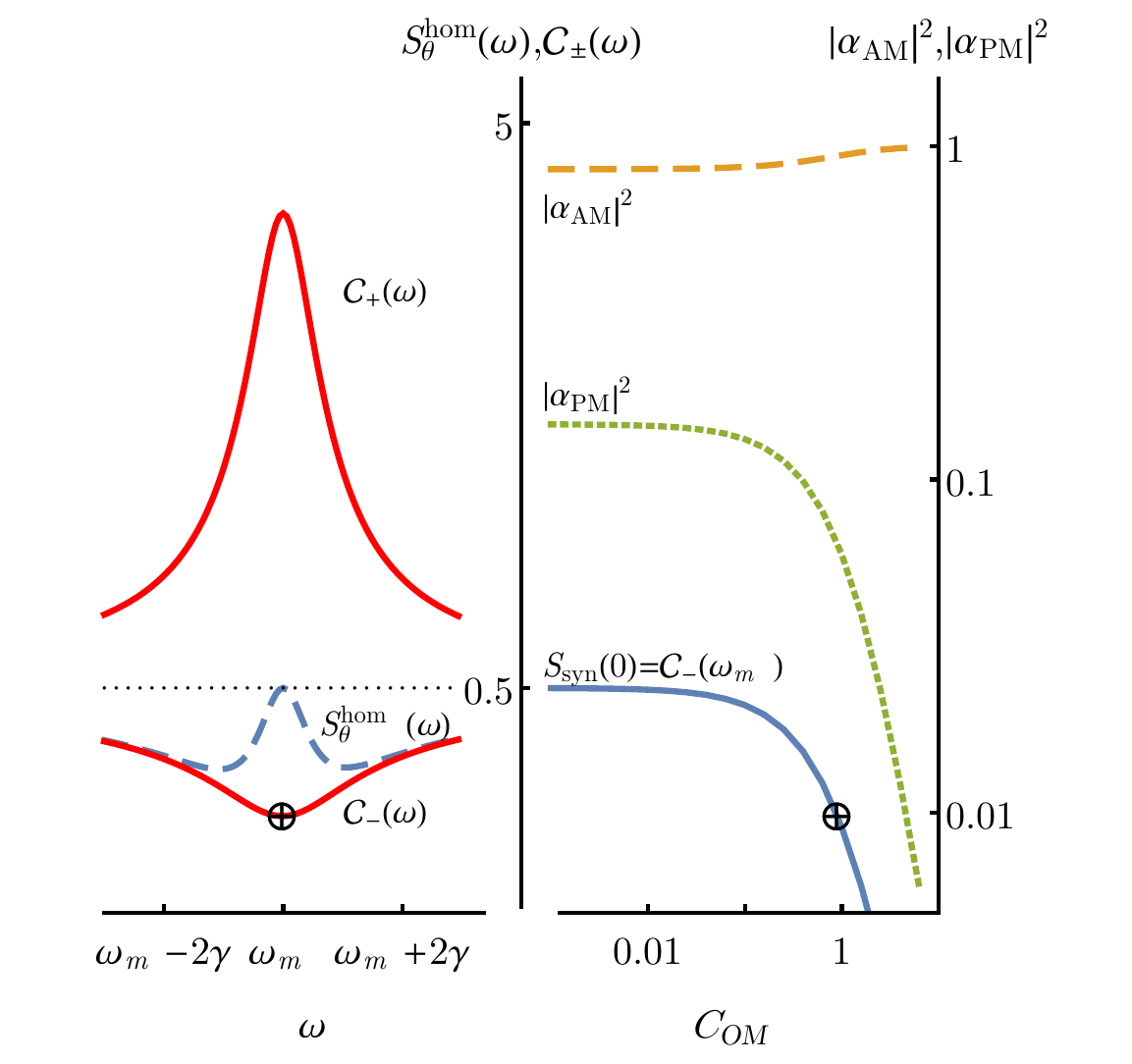}
\caption{(Color online) Synodyne detection yields noise below the vacuum level due to complex squeezing. (a) Eigenvalues of the covariance matrix as a function of frequency (red, solid) and spectral density of a homodyne measurement (blue, dashed) at $C_{OM}=0.9$. The dotted black line marks the vacuum noise level 1/2. (b) Noise spectral density of a synodyne measurement at $\omega=0$ and minimal eigenvalue of the covariance matrix (solid line, left axis). Relative power of $|\alpha_{\mathrm{AM}}|^2$ (orange, dashed) and $|\alpha_{\mathrm{PM}}|^2$ (green, dotted) of the LO tones, right axis. Powers and phases of the LO tones are chosen to minimize the detected noise. The marker indicates points of equal cooperativity in (a) and (b). System parameters are $\omega_m/\kappa=0.2$, $\gamma_m/\kappa=0.002$, $\bar{\nu}=0$.}
\label{fig2}
\end{figure}

However, at $\omega=\omega_m$, where $\chi_m$, and hence force sensitivity, has its highest magnitude, the mechanical susceptibility and therefore also $\left[\mathcal{C}(\omega_m)\right]_{12}$, is purely imaginary.  The large mechanical susceptibility leads to strong correlations in the cavity output light, as seen from the eigenvalues $\mathcal{C}_\pm$ of $\mathcal{C}(\omega)$, shown in Fig.\ \ref{fig2}(a).  These correlations are invisible to homodyne detection, which shows no squeezing at the mechanical resonance frequency.  This absence prevents ponderomotive squeezing from improving the sensitivity of a homodyne-based optomechanical force sensor below the SQL on mechanical resonance~\cite{linamp, freqdephomodyne}.

To measure and exploit all correlations, it is necessary to devise a measurement procedure more adapted to the signal being measured. Therefore, we consider a more general LO waveform, $\alpha(t) e^{-i \omega_0 t}$, which, after being mixed with the signal field and measured on a square-law detector, measures the operator
\begin{align}
\hat{\Xi}(t)=\mathrm{Re}(\alpha(t))\hat{X}_{\mathrm{AM}}(t)+\mathrm{Im}(\alpha(t))\hat{X}_{\mathrm{PM}}(t).
\end{align}
We seek a detector that suitably combines information obtained in different measurement quadratures during different temporal phases, e.g.\ times selected by either the $\cos(\omega_s t)$ or $\sin(\omega_s t)$ functions, where $\omega_s$ can be chosen freely. Consider an LO that is a sum of two tones, at the optical frequencies $\omega_0 \pm \omega_s$, for which $\alpha(t)= \alpha_- e^{i \omega_s t} + \alpha_+ e^{-i\omega_s t}$. Bi-chromatic local oscillators have been proposed for the detection of entanglement between states with different frequencies~\cite{boyd} and the operation of interferometers~\cite{buonanno} and have been demonstrated experimentally~\cite{LiYuZhang}.

 The additional tone in the LO provides the freedom to match the amplitudes and phases of $\alpha_\mathrm{PM}$ with the correlations encoded in the the Stokes and anti-Stokes mechanical sidebands that result from the mechanical response; in light of this matching, we call the resulting optical receiver a \emph{synodyne detector}. Its measurement operator in the frequency domain reads 
\begin{align}
\tilde{\Xi}(\omega)=
\frac{1}{\sqrt{2}}&\left[\alpha_\mathrm{AM}\tilde{X}_\mathrm{AM}^\mathrm{out}(\omega-\omega_s)
+\alpha_\mathrm{AM}^*\tilde{X}_\mathrm{AM}^\mathrm{out}(\omega+\omega_s)
\right.
\nonumber\\
&
\left.+\alpha_\mathrm{PM}\tilde{X}_\mathrm{PM}^\mathrm{out}(\omega+\omega_s)
+\alpha_\mathrm{PM}^*\tilde{X}_\mathrm{PM}^\mathrm{out}(\omega-\omega_s)
\right],\label{synodyneoperator}
\end{align}
with \mbox{$\alpha_{\mathrm{AM}}=\frac{\alpha_+ + \alpha_{-}^*}{\sqrt{2I}}$}, $\alpha_{\mathrm{PM}}=i\frac{\alpha_{+}^*-\alpha_{-}}{\sqrt{2I}}$ and $I=|\alpha_+|^2+|\alpha_-|^2$.
For $\omega_s=\omega_m$, the two sidebands overlap at the dc part of the signal and the noise spectral density of the detector output is
\begin{align}\label{twotonesynodynepower}
S_\mathrm{syn}(0)=&|\alpha_{\mathrm{AM}}|^2[\mathcal{C}(\omega_m)]_{11}
+|\alpha_{\mathrm{PM}}|^2[\mathcal{C}(\omega_m)]_{22}
\nonumber\\
&+2\mathrm{Re}\left(\alpha_{\mathrm{AM}}^*\alpha_{\mathrm{PM}}^*[\mathcal{C}(\omega_m)]_{12}\right),
\end{align}
  Comparing Eqs.(\ref{homodynepsd}) and (\ref{twotonesynodynepower}) reveals that, unlike the homodyne measurement, the synodyne setting can exploit the full magnitude of the complex-valued correlation term to reduce the noise in the detected signal. This cancellation occurs only in a narrow frequency band of width $~\gamma_m$ around dc and may be seen as an extension of the benefits of variational read-out to the mechanical resonance~\cite{nir}, although its practical implementation is qualitatively different~\cite{freqdephomodyne}.

  Comparing Eqs.(\ref{homodynepsd}) and (\ref{twotonesynodynepower}) reveals that, unlike the homodyne measurement, the synodyne setting can exploit the full magnitude of the complex-valued correlation term to reduce the noise in the detected signal. This cancellation occurs only in a narrow frequency band of width $~\gamma_m$ around dc and may be seen as an extension of the benefits of variational read-out to the mechanical resonance~\cite{nir}, although its practical implementation is qualitatively different~\cite{freqdephomodyne}.

Complex correlations between AM and PM fluctuations can in principle also be detected by heterodyne detection. In that case, the offset of the LO frequencies from the carrier frequency $\omega_0$ is much larger than the signal frequencies of interest. However, the heterodyne receiver admits additional vacuum noise from frequencies that do not contain any signal~\cite{buonanno}. This additional noise prevents the heterodyne signal from reaching noise levels below that of homodyne detection, even after accounting for complex correlations. In contrast, synodyne detection does not lead to additional noise.

The optomechanical response at $\omega_m$ also appears within the synodyne detection record at the frequency $\omega_s+\omega_m=2\omega_m$, see Eq. (\ref{synodyneoperator}). However, in the limit of strong complex correlations ($G\gg\sqrt{\kappa\gamma}$), these extra ac signals contain a diminishing amount of information on the noise operator measured by the dc signal and we neglect it in the following. 

As shown in Fig.\ \ref{fig2}(b), an optimal choice of $\alpha_\mathrm{PM}$ at each value of the optomechanical cooperativity $C_{OM} = 2G^2 / (\kappa \gamma_m)$ results in the detected noise matching the lower eigenvalue $\mathcal{C}_-$ of the covariance matrix.  The process can be tailored to reveal correlations irrespective of their complex phase and results in a noise level lower than that detected when the input port is blocked; the detected signal is truly squeezed. 
This squeezing can be created for any temperature of the mechanical oscillator, because back-action overpowers the transduction of thermal mechanical noise at sufficiently high cooperativity (compare Eqs. (\ref{transduction}) and (\ref{backaction}), see also~\cite{TeufelBA} and ~\cite{supp}).

The optimal values of $\alpha_+$ and $\alpha_-$ chosen to reduce detected noise are reminiscent of the mechanical sideband asymmetry observed at the output of a driven cavity optomechanical system \cite{amirzpt,E3zpt}.  Indeed, the sideband asymmetry observable with standard detection techniques reflects exactly the unequal-time correlations discussed in this work \cite{vitali, E3zpt, amirzpt,khalili,schwabzpt}.

\begin{figure}[t]
\begin{tabular}{r}
\includegraphics[width=0.39\textwidth]{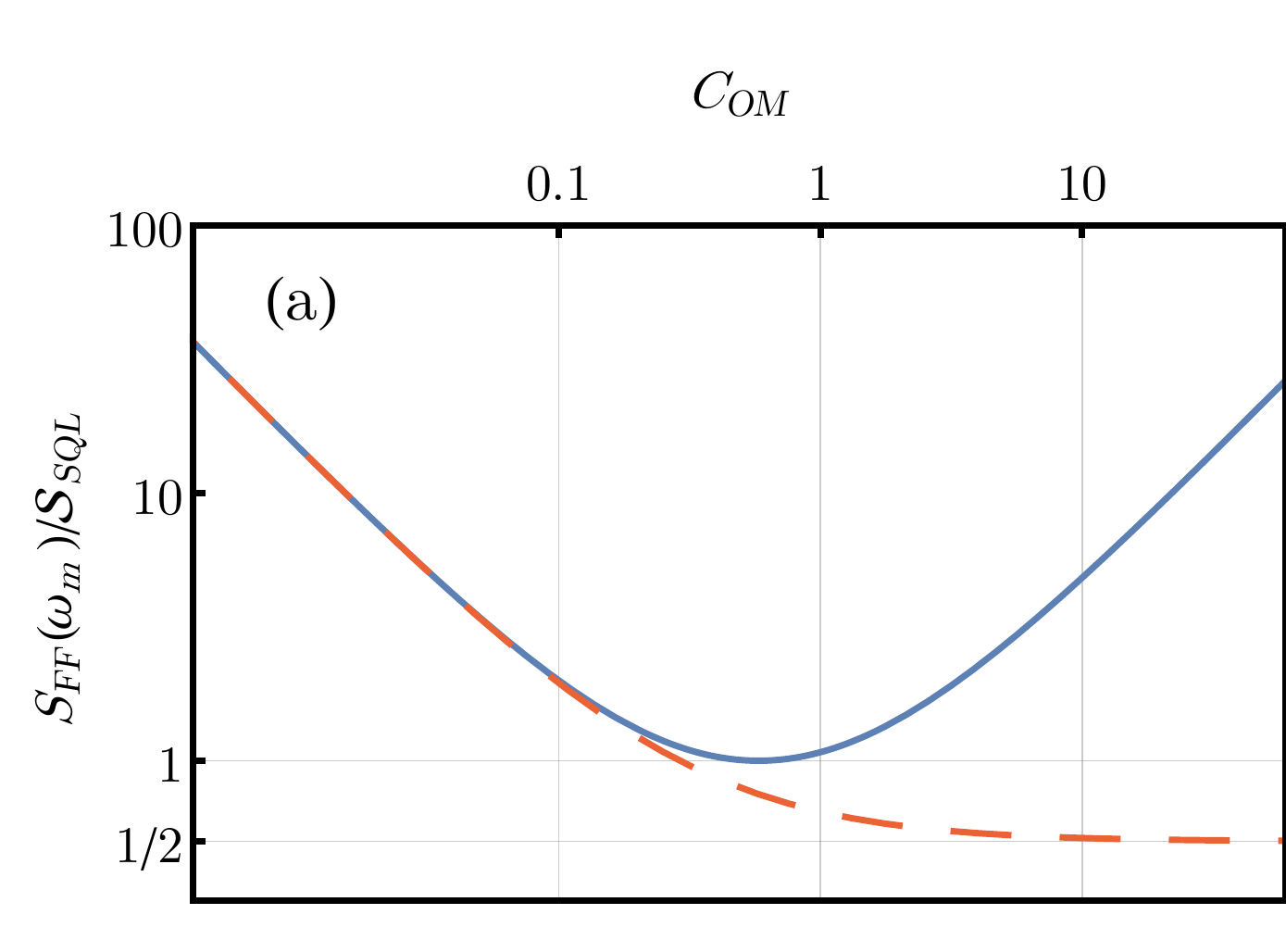}
\\
\includegraphics[width=0.38\textwidth]{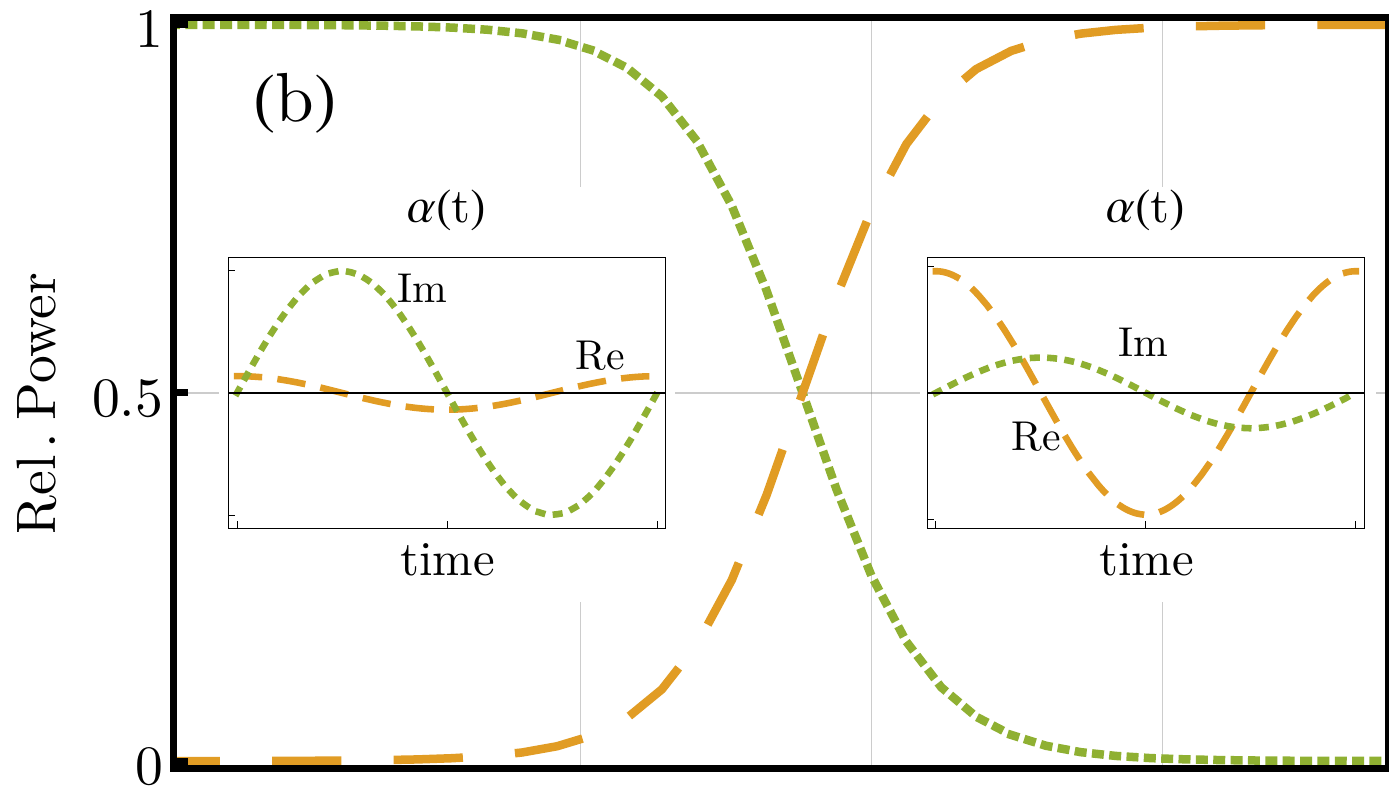}
\end{tabular}
\caption{(Color online) Exploiting complex squeezing in an optomechanical force sensor. (a) Measurement imprecision of homodyne (blue, solid) and synodyne measurements of a single component (orange, dashed) of an external force as a function of cooperativity. Both curves are normalized to the value of a homodyne measurement at the SQL. (b) Relative powers $|\alpha_{\mathrm{PM}}|^2$ (green, dotted) and $|\alpha_{\mathrm{AM}}|^2$  (yellow, dashed) in a synodyne measurement as a function of cooperativity. The two insets show the measurement strengths $\mathrm{Re}(\alpha(t))$ (yellow, dashed) and $\mathrm{Im}(\alpha(t))$ (green, dotted) for a single cycle of the mechanical oscillator at $C_{OM}=0.08$ (left) and $2$ (right). Remaining parameters same as Fig.~\ref{fig2}.}
\label{fig3}
\end{figure}

Now, we consider the utility of complex squeezing and synodyne detection to improve the performance of an optomechanical force sensor.  Instead of choosing $\alpha_\mathrm{PM}$ to minimize the detector noise level, one aims to minimize force measurement imprecision. A synodyne detector is sensitive to the temporal phase of an external force via the choice of the phases of the LO tones $\alpha_{\mathrm{PM}}$. Let us consider the temporal phase of detection to be aligned with that of the external force. For an external force with known temporal phase, the SQL can be defined as the minimal imprecision of a single-component homodyne measurement of the phase quadrature and given by $\mathcal{S}_\mathrm{SQL}=\frac{[\mathcal{C}(\omega_m)]_{22}}{2|T_p(\omega_m)|^2}$. For a synodyne measurement, the imprecision of the dc signal is given by $S_{FF} = \frac{S_\mathrm{syn}(0)}{2|\alpha_\mathrm{PM}|^2|T_p(\omega_m)|^2}$, where $\alpha_\mathrm{PM}$ has to be optimized for every cooperativity. The noise $[\mathcal{C}(\omega_m)]_{22}$ contains optical shot-noise and back-action, see Eq.~(\ref{pmoutput}). In a synodyne measurement, the back-action contribution can be cancelled by the complex correlations, the last term in Eq.~(\ref{twotonesynodynepower}). The shot-noise contribution becomes negligable for large input powers. These two effects lead to a measurement sensitivity that monotonically increases with $C_{OM}$ and is only limited by noise from the mechanical oscillator. This behavior is evident in the imprecision curves of a homodyne and synodyne measurement plotted in Fig.~\ref{fig3}(a).

The imprecision curve of a synodyne receiver is similar to that of back-action evading detection schemes, in which the cavity is driven with light that is modulated at twice the frequency of motion, so that measurement back-action accumulates only in the unmeasured quadrature~\cite{bae, mechsqueezing1,mechsqueezing2}.  Such a scheme is susceptible to parametric instabilities caused by the modulated radiation pressure force applied to the oscillator~\cite{paraminst}. Additionally, it is limited to measurements on systems satisfying $\omega_m/\kappa\gg 1$. With our proposed method, which may be called a \emph{back-action accounting} scheme, the cavity is driven with a steady cavity input, avoiding parametric instabilities. The synodyne receiver measures the radiation-pressure force noise visible in the AM fluctuations of the cavity output, and subtracts that noise from the PM fluctuations of the light that occur after a quarter-cycle of mechanical motion. To properly account for back-action noise, the relative strength with which the two types of fluctuations are detected varies with optomechanical cooperativity $C_{OM}$, as detailed in Fig.\ \ref{fig3}(b). The benefits of back-action accounting are available for any value of the sideband resolution $\omega_m/\kappa$, (Figs.~\ref{fig2} and \ref{fig3}).

To summarize, we have broadened the definition of inhomogeneous squeezing to include complex squeezing, which accounts for unequal-time correlations between AM and PM fluctuations. We have restricted our discussion to stationary noise processes, where auto-correlations do not exhibit temporal phases. For a non-stationary process, one would consider instead the full $4\times 4$ covariance matrix between operators defined in Eq.~(\ref{weightedops}). The generalization of our treatment to such situations is straightforward.

We introduced synodyne detection as a means to detect one temporal phase of the squeezed spectrum at a specific detection frequency $\omega_s$.  Synodyne detection allows one to freely choose the measurement basis in this four-dimensional space of a non-stationary covariance matrix and, for instance, to reveal any complex phase of the covariance $[\mathcal{C}(\omega)]_{12}$. For the example of the output of an optomechanical cavity, synodyne detection allows full back-action accounting for phase-sensitive force detection at any frequency across the ponderomotive squeezing spectrum. We have not addressed if such a detector is optimal in terms of sensitivity, bandwidth, or adaptation to detecting forces at frequencies away from the mechanical resonance. To answer such questions, it may be interesting to consider a broader selection of synodyne detection wave forms, which might involve measuring different external force components at different times, measure particular temporal force profiles or generalize back-action accounting over a broader bandwidth.

\begin{acknowledgments}
This work was supported by the Air Force Office of Scientific Research, the Villum Foundation Center of Excellence,  and NSF. N.S.\ was
supported by a Marie Curie International Outgoing Fellowship, J.K.\ and S.S.\ by the US
Department of Defense through the National Defense Science and Engineering Graduate
Fellowship Program, and L.F.B.\ by the Swiss National Science Foundation. L.F.B.\ acknowledges helpful discussions with \mbox{K.~M{\o}lmer}, \mbox{D.~Petrosyan} and \mbox{N.~Kampel.}
\end{acknowledgments}

\end{document}